\documentstyle[12pt]{article}
\begin{document}
\parskip 10pt plus 1pt
\title{GENERALISING THE CONCEPT OF MAXIMAL SYMMETRY OF SPACETIMES
IN THE PRESENCE OF TORSION}
\author{
{\it $^a$ Debashis Gangopadhyay and $^b$ Soumitra Sengupta}
\\{$^a$ S.N.Bose National Centre For Basic Sciences}\\
{DB-17, Sector-I, Salt Lake, Calcutta-700064, INDIA}\\
{$^b$ Department of Physics, Presidency College}\\
{86/1, College Street , Calcutta-700073, INDIA}
}
\baselineskip=10pt
\date{}
\maketitle
\begin{abstract}
A possible generalisation is given to the meaning of maximal symmetry in
the presence of torsion.

PACS NO: 04.20.cv ; 02.40.ky ; 11.27 + d
\end{abstract}

\newpage
In $N$ dimensions, a metric that admits the maximum number $N(N + 1)/2$ of
Killing vectors is said to be maximally symmetric.A maximally symmetric space
is homogeneous and isotropic about all points [1].Such spaces are
of natural interest in the general theory of relativity as they correspond to
spaces of globally constant curvature which in turn is related to the concepts
of homogeneity and isotropy.We propose a possible generalisation of the
meaning of maximal symmetry when torsion is present.The motivation stems from
string theory where there is a second rank antisymmetric tensor field in the
spectrum that is identified with the background torsion.Therefore a maximally
symmetric solution of the classical string equations of motion necessarily
requires a generalisation of the definition itself.

The presence of torsion implies that the affine connections
$\bar{\Gamma}^{\alpha}_{\mu\nu}$ are asymmetric and contain an antisymmetric
part $ H ^{\alpha}_{\mu\nu}$ in addition to the usual symmetric term
$\Gamma^{\alpha}_{\mu\nu}$ [2,3]:
$$\bar{\Gamma}^{\alpha}_{\mu\nu} = \Gamma^{\alpha}_{\mu\nu} +
H^{\alpha}_{\mu\nu}\eqno(1)$$
The notation has been chosen keeping string theory in mind where the
antisymmetric third rank tensor $H_{\alpha\mu\nu} = \partial_{(\alpha}
B_{\mu\nu)}$
is identified with the background torsion and $B_{\mu\nu}$ is the second rank
antisymmetric tensor field mentioned above.Note that $H^{\alpha}_{\mu\nu}$ are
completely arbitrary to start with.

Defining covariant derivatives with respect to $\bar{\Gamma}^{\alpha}_{\mu\nu}$
we have for a vector field $V_{\mu}$ :
$$V_{\mu;\nu;\beta} - V_{\mu;\beta;\nu} = -\bar{R}^{\lambda}_{\mu\nu\beta}
V_{\lambda} + 2 H^{\alpha}_{\beta\nu} V_{\mu;\alpha} \eqno(2)$$
where
$$\bar{R}^{\lambda}_{\mu\nu\beta} = R^{\lambda}_{\mu\nu\beta} +
\tilde{R}^{\lambda}_{\mu\nu\beta} \eqno(3a)$$
$$R^{\lambda}_{\mu\nu\beta} = \Gamma^{\lambda}_{\mu\nu, \beta} -
\Gamma^{\lambda}_{\mu\beta, \nu} +
\Gamma^{\alpha}_{\mu\nu}\Gamma^{\lambda}_{\alpha\beta} -
\Gamma^{\alpha}_{\mu\beta}\Gamma^{\lambda}_{\alpha\nu} \eqno(3b)$$
$$\tilde{R}^{\lambda}_{\mu\nu\beta} = H^{\lambda}_{\mu\nu, \beta} -
H^{\lambda}_{\mu\beta, \nu} +
H^{\alpha}_{\mu\nu}H^{\lambda}_{\alpha\beta} -
H^{\alpha}_{\mu\beta}H^{\lambda}_{\alpha\nu}$$
$$\phantom{\tilde{R}^{\lambda}_{\mu\nu\beta}} +
\Gamma^{\alpha}_{\mu\nu}H^{\lambda}_{\alpha\beta} -
H^{\alpha}_{\mu\beta}\Gamma^{\lambda}_{\alpha\nu} +
H^{\alpha}_{\mu\nu}\Gamma^{\lambda}_{\alpha\beta} -
\Gamma^{\alpha}_{\mu\beta}H^{\lambda}_{\alpha\nu} \eqno(3c) $$
The generalised curvature $\bar{R}^{\lambda}_{\mu\nu\beta}$ does not have the
usual symmetry (antisymmetry) properties.Note that the last term on the
right hand side $(2)$ is obviously a tensor.Hence
$\bar{R}^{\lambda}_{\mu\nu\beta}$ is also a tensor.

Now, the determination of all infinitesimal isometries of a metric is
equivalent to determining all Killing vectors $\xi_{\mu}$ of the metric.A
Killing vector is defined through the Killing condition:
$$\xi_{\mu\enskip ;\enskip\nu} + \xi_{\nu\enskip ;\enskip\mu} = 0\eqno(4)$$
and it is easily verified that this condition is preserved also in the
presence of the torsion $H^{\alpha}_{\mu\nu}$.

Equation $(2)$ for a Killing vector hence takes the form:
$$\xi_{\mu;\nu;\beta} - \xi_{\mu;\beta;\nu} = -
\bar{R}^{\lambda}_{\mu\nu\beta} \xi_{\lambda} +
2 H^{\alpha}_{\beta\nu}  \xi_{\mu;\alpha}\eqno(5)$$
The $H^{\alpha}_{\beta\nu}$ are arbitrary and so we may choose them to be
such that
$$H^{\alpha}_{\beta\nu}  \xi_{\mu ;\alpha} = 0 \eqno(6)$$
This is a constraint on the $H^{\alpha}_{\beta\nu}$ and not the Killing
vectors  $\xi_{\mu}$.The commutator of two covariant derivatives of a
Killing vector thus becomes:
$$\xi_{\mu;\nu;\beta} - \xi_{\mu;\beta;\nu} = -
\bar{R}^{\lambda}_{\mu\nu\beta} \xi_{\lambda}\eqno(7)$$

We now impose the cyclic sum rule on $\bar{R}^{\lambda}_{\mu\nu\beta}$:
$$\bar{R}^{\lambda}_{\mu\nu\beta} + \bar{R}^{\lambda}_{\nu\beta\mu} +
\bar{R}^{\lambda}_{\beta\mu\nu} = 0 \eqno(8a)$$
As $R^{\lambda}_{\mu\nu\beta} +R^{\lambda}_{\nu\beta\mu} +
R^{\lambda}_{\beta\mu\nu} = 0$  the constraint $(8a)$ implies
$$\tilde{R}^{\lambda}_{\mu\nu\beta} + \tilde{R}^{\lambda}_{\nu\beta\mu} +
\tilde{R}^{\lambda}_{\beta\mu\nu} = 0 $$
and this reduces to
$$ H^{\lambda}_{\mu\nu,\beta} +
H^{\alpha}_{\mu\nu}\bar{\Gamma}^{\lambda}_{\alpha\beta} +
H^{\lambda}_{\nu\beta,\mu} +
H^{\alpha}_{\nu\beta}\bar{\Gamma}^{\lambda}_{\alpha\mu} +
H^{\alpha}_{\beta\mu,\nu} +
H^{\alpha}_{\beta\mu}\bar{\Gamma}^{\lambda}_{\alpha\nu} = 0\eqno(8b) $$

Hence adding $(7)$ and its two cyclic permutations and using $(4)$, the
relation $(7)$ can be recast into
$$\xi_{\mu;\nu;\beta} = -
\bar{R}^{\lambda}_{\beta\nu\mu}\xi_{\lambda}\eqno(9)$$

Therefore given $\xi_{\lambda}$ and $\xi_{\lambda;\nu}$ at some point $X$ ,
we can determine the second derivatives of $\xi_{\lambda}$$(x)$ at $X$ from
$(9)$.Then following the usual arguments [1], any Killing vector
$\xi^{n}_{\mu}$$(x)$ of the metric $g_{\mu\nu}$$(x)$ can be expressed as
$$\xi^{n}_{\mu}(x) = A^{\lambda}_{\mu}(x\enskip ;\enskip X)
\xi^{n}_{\lambda}(X) + C^{\lambda\nu}_{\mu} (x\enskip;\enskip X)
\xi^{n}_{\lambda ;\nu}(X)\eqno(10)$$
where $A^{\lambda}_{\mu}$ and $C^{\lambda\nu}_{\mu}$ are functions that
depend on the metric, {\it torsion} and $X$ but not on the initial values
$\xi_{\lambda}$$(X)$ and $\xi_{\lambda;\nu}$$(X)$, and hence are the same
for all Killing vectors.Also note that the torsion fields present in
$A^{\lambda}_{\mu}$$(x ; X)$ and  $C^{\lambda\nu}_{\mu}$$(x ; X)$ obey the
constraint $(6)$.A set of Killing vectors $\xi^{n}_{\mu}$$(x)$  is said to
be independent if they do not satisfy any relation of the form
$\Sigma_{n} d_{n} \xi^{n}_{\mu}$$(x)$ $= 0$,with constant coefficients
$d_{n}$.It therefore follows that there can be at most $N(N+1)/2$
independent Killing vectors in $N$ dimensions,{\it even in the presence of
torsion provided the torsion fields satisfy the constraints (6) and (8b)}.

Now, $\bar{R}_{\lambda\mu\nu\beta}$ is antisymmetric in the indices
$(\lambda, \mu)$  and $(\nu, \beta)$.This follows from the fact that
$R_{\lambda\mu\nu\beta}$  and $\tilde{R}_{\lambda\mu\nu\beta}$ both have
these properties.This can be verified through an elaborate but
straightforward calculation.Then proceeding as in [1],we have
(using -$\bar{R}^{\alpha}_{\mu\nu\alpha} = \bar{R}_{\mu\nu}$ etc.)
$$(N-1) \bar{R}_{\lambda\mu\nu\beta} = \bar{R}_{\beta\mu}g_{\lambda\nu} -
\bar{R}_{\nu\mu}g_{\lambda\beta}$$
i.e.$$(N-1) R_{\lambda\mu\nu\beta} + (N-1) \tilde{R}_{\lambda\mu\nu\beta}$$
$$= R_{\beta\mu} g_{\lambda\nu} - R_{\nu\mu} g_{\lambda\beta} +
\tilde{R}_{\beta\mu} g_{\lambda\nu} -
\tilde{R}_{\nu\mu} g_{\lambda\beta}\eqno(11)$$
where $N$ is the number of dimensions.
$R_{\lambda\mu\nu\beta}$, $R_{\beta\mu}$ are functions of the symmetric affine
connections $\Gamma$ only, whereas $\tilde{R}_{\lambda\mu\nu\beta}$,
$\tilde{R}_{\beta\mu}$ are functions of both $\Gamma$ and $H$.Moreover,
$\Gamma$ and $H$  are independent to start with.Hence in equation $(11)$ we
are justified in equating corresponding terms on both sides to get the
following two equations:
$$(N-1) R_{\lambda\mu\nu\beta} = R_{\beta\mu} g_{\lambda\nu} -
R_{\nu\mu} g_{\lambda\beta}\eqno(12a)$$
$$(N-1) \tilde{R}_{\lambda\mu\nu\beta} = \tilde{R}_{\beta\mu} g_{\lambda\nu} -
\tilde{R}_{\nu\mu} g_{\lambda\beta}\eqno(12b)$$

The above two equations lead to
$$R_{\lambda\mu\nu\beta} = R^{\alpha}_{\alpha}(g_{\lambda\nu} g_{\mu\beta} -
g_{\lambda\beta} g_{\nu\mu})/ N(N - 1)\eqno(13a)$$
and
$$\tilde{R}_{\lambda\mu\nu\beta} = \tilde{R}^{\alpha}_{\alpha}
(g_{\lambda\nu} g_{\mu\beta} - g_{\lambda\beta} g_{\nu\mu}/ N(N -1)\eqno(13b)$$

It is appropriate to note here that
$$\tilde{R}^{\alpha}_{\alpha\nu\beta} = 0 \eqno(14a)$$
$$\tilde{R}^{\alpha}_{\mu\nu\alpha} = H^{\alpha}_{\mu\nu,\alpha} -
H^{\alpha}_{\mu\gamma} H^{\gamma}_{\alpha\nu} +
H^{\alpha}_{\mu\nu} \Gamma^{\gamma}_{\alpha\gamma} -
H^{\alpha}_{\gamma\nu} \Gamma^{\gamma}_{\mu\alpha} -
H^{\alpha}_{\mu\gamma} \Gamma^{\gamma}_{\alpha\nu} \eqno(14b)$$

If we now demand that $\tilde{R}_{\mu\nu} = \tilde{R}_{\nu\mu}$ we have the
constraint
$$H^{\alpha}_{\mu\nu,\alpha} +
H^{\alpha}_{\mu\nu} \Gamma^{\gamma}_{\alpha\gamma} +
H^{\alpha}_{\nu\gamma} \Gamma^{\gamma}_{\alpha\mu} -
H^{\alpha}_{\mu\gamma}  \Gamma^{\gamma}_{\alpha\nu} = 0\eqno(15)$$
which is the same as $(8b)$ with $\beta=\lambda$.Hence $(8b)$ implies that
$\tilde{R}_{\mu\nu}$ is symmetric.

Under these circumstances we have
$$R_{\mu\nu} = (1/N) g_{\mu\nu}R^{\alpha}_{\alpha}\eqno(16a)$$
$$\tilde{R}_{\mu\nu} = (1/N) g_{\mu\nu}\tilde{R}^{\alpha}_{\alpha}
= H^{\alpha}_{\mu\beta} H^{\beta}_{\alpha\nu}\eqno(16b)$$

Now using arguments similar to those given in Ref.[1] for the Bianchi
identities we can conclude that in the presence of torsion fields
satisfying constraints discussed before :
$$\bar{R}_{\lambda\mu\nu\beta} =
R_{\lambda\mu\nu\beta} + \tilde{R}_{\lambda\mu\nu\beta} =
(K+ \tilde{K})(g_{\lambda\nu}g_{\mu\beta} - g_{\lambda\beta}g_{\mu\nu}) =
\bar{K}(g_{\lambda\nu}g_{\mu\beta} - g_{\lambda\beta}g_{\mu\nu})\eqno(17)$$
where
$$R^{\alpha}_{\alpha} = constant = K N (1 - N)\eqno(18a)$$
$$\tilde{R}^{\alpha}_{\alpha} = H^{\mu\lambda}_{\beta} H^{\beta}_{\lambda\mu}=
constant = \tilde{K} N (1 - N)\eqno(18b)$$
$$\bar{K} = K + \tilde{K} = constant\eqno(18c)$$
(In deriving the above results from  the Bianchi identities we have used
the fact that for a flat metric the curvature constant $K=0$.Hence demanding
$\bar{K}=0$ for a (globally) zero curvature space means that $\tilde{K}=0$
which in turn means that the torsion must vanish.)

Therefore, in the presence of torsion the criteria of maximal symmetry has
been generalised through the equations $(13b)$, $(17)$ and $(18)$.The physical
meaning is still that of a globally constant curvature (which now also has a
contribution coming from the torsion).The torsion fields are, however,
subject to the constraints embodied in equations $(6)$, $(8b)$ and $(18b)$
and these constraints are mutually consistent.Hence our usual concepts of
the  homogeneity and isotropy of space have a more generalised footing in
the presence of torsion.The thing to note is that the existence of torsion
does not necessarily jeopardise the prevalent concepts of an isotropic and
homogeneous spacetime.All that is required is that the torsion fields obey
certain mutually consistent constraints.These constraints, in some sense,
ensure that the usual physical meaning of an isotropic and homogeneous
spacetime remains intact.

It is also worth mentioning here  the relevance of the present work in the
context of string theory.The low energy string effective action posseses
for time dependent metric $G_{\mu\nu} (\mu,\nu = 1,2,....d)$, torsion
$B_{\mu\nu}$ and dilaton $\phi$ background fields a full continuous
$O(d,d)$ symmetry under which "cosmological" solutions of the equations of
motion are transformed into other inequivalent solutions [4,5].Subsequently,
it was shown that given a classical background in string theory
independent of $d$ of the space-time coordinates , other classical
backgrounds are possible by $O(d)\otimes O(d)$ transformations on the
solution [6,7].A solution with zero torsion is necessarily connected to a
solution with non-zero torsion.Therefore, in the framework of the
generalised maximal symmetry discussed here it is worth studying whether this
generalised maximal symmetry is preserved under the $O(d)\otimes O(d)$
transformation.The results of this investigation will be reported elsewhere.
We thank A.K.Raychaudhuri for illuminating discussions.One of us
(S.S.G.) gratefully acknowledges discussions with G.Ellis.

\end{document}